\documentclass{moriond}

\usepackage{slashed}
\usepackage{amsmath}

\def\be{\begin{equation}}
\def\ee{\end{equation}}
\def\bea{\begin{eqnarray}}
\def\eea{\end{eqnarray}}

\newcommand{\Photo}{\includegraphics[height=35mm]{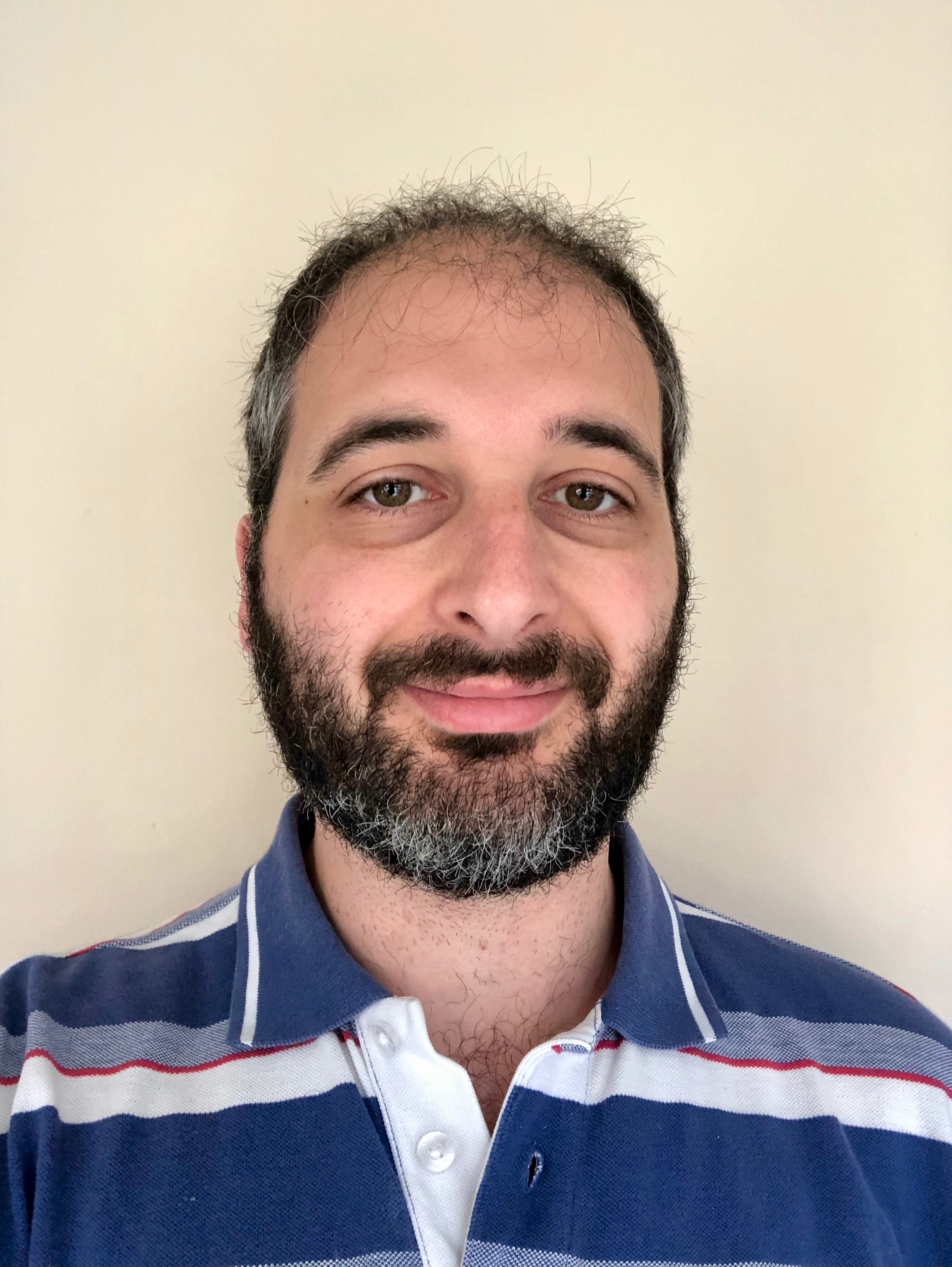}}

\begin{document}
\begin{flushright}
CP3-19-32	
\end{flushright}
\vspace*{4cm}
\title{TESTABILITY OF LEPTOGENESIS WITH THREE RH-NEUTRINOS BELOW THE ELECTROWEAK SCALE}

\author{MICHELE LUCENTE}

\address{Centre for Cosmology, Particle Physics and Phenomenology (CP3)
Universit\'e catholique de Louvain
Chemin du Cyclotron, 2
B-1348 Louvain-la-Neuve,
Belgium}

\maketitle\abstracts{The Standard Model extended with right-handed neutrinos whose masses are below the electroweak scale provides a simultaneous solution for the origin of neutrino masses and of the baryon asymmetry of the Universe, that can be tested in current experiments. If three right-handed neutrinos participate to the processes, their parameter space of solutions extends to very large mixing angles, saturating the current experimental constraints. Solutions with right-handed neutrino masses at the GeV scale can be probed in the decay of $B$ mesons at the LHC. For this channel the collision of isotopes of intermediate mass such as Ar provides a better sensitivity per unit of running time compared to collisions with protons.}

\section{Introduction}
Among the hints for the existence of new - yet undiscovered - physics, there are two seemingly unrelated observations that cannot be accounted for in the Standard Model (SM) of particle physics: i) neutrinos are massive (and leptons mix) and ii) the baryon asymmetry of the Universe (BAU),  defined as the difference between the number densities of baryons and anti-baryons normalised to the photon number density, has the value~\cite{Aghanim:2018eyx} $\eta_{\Delta B}  = \left(n_b - n_{\overline{b}}\right)/n_\gamma = \left( 6.13 \pm 0.03 \right) \times 10^{-10}$. At the same time, by looking at the field content of the SM, one can realise that there is a certain asymmetry in it: all the fermionic fields appear in both states of chirality, left- (LH) and right-handed (RH), with the exception of neutrinos, which only possess a LH component. It is thus natural to ask what are the phenomenological consequences of extending the SM field content by including also RH-neutrino fields in it, and study if this minimal extension can address the above mentioned observational problems of the SM.

\subsection{Neutrino masses and leptogenesis}
By including a number $n$ of RH-neutrino fields $N_I$ in the SM field content, the Lagrangian is extended by the inclusion of the following most-general gauge and Lorentz invariant renormalizable operators:
\be\label{eq:Lagrangian}
\mathcal{L} = \mathcal{L}_\text{SM} + i \overline{N_I} \slashed{\partial} N_I - \left( F_{\alpha I} \overline{\ell_L^{\alpha}} \varepsilon \phi^* N_I + \frac{M_{IJ}}{2} \overline{N_I^c} N_J + h.c. \right),
\ee 
where $\ell^\alpha_L$ are the LH SU(2) lepton doublets ($\alpha = e,\mu,\tau$), $\phi$ is the Higgs doublet, $\varepsilon$ the total antisymmetric bi-dimensional tensor. $F_{\alpha I}$ are dimensionless Yukawa couplings, while $M$ is an $n$-dimensional symmetric matrix of Majorana mass terms. After electroweak symmetry breaking (EWSB) the Higgs field develops a vacuum expectation value, $<\phi>\ = v \simeq 174$ GeV, and the Lagrangian~(\ref{eq:Lagrangian}) generates a non-vanishing Majorana mass matrix $m_\nu$ for the left-handed neutrinos,
\be\label{eq:light_masses}
m_\nu = -\Theta M \Theta^T + \mathcal{O}\left(\Theta^3\right) \simeq - v^2 F M^{-1} F^T.
\ee 
Here $\Theta = v F M^{-1}$ is the Seesaw expansion parameter, which must have perturbative values because of the phenomenological constraints~\cite{Blennow:2016jkn} on the deviation from unitarity of the leptonic PMNS mixing matrix, $N_{\textrm{PMNS}}  = \left(1 - \Theta \Theta^\dagger / 2 \right)U_\nu + \mathcal{O}\left(\Theta^3\right)$; $U_\nu$ is the unitary matrix that diagonalises $m_\nu$ and $N_{\textrm{PMNS}}$ the physical mixing matrix measured in neutrino oscillation experiments.

It is remarkable that the Lagrangian~(\ref{eq:Lagrangian}) provides, at the same time, all the necessary ingredients~\cite{Sakharov:1967dj} for a successful baryogenesis via leptogenesis~\cite{Fukugita:1986hr}: the Yukawa matrix $F$ is in general complex, providing a new source of $CP$-violation, while the new degrees of freedom $N_I$ can deviate from thermal equilibrium during their evolution in the early Universe. These features allow to generate a lepton asymmetry, which is subsequently partially converted into a non-vanishing baryon asymmetry by non-perturbative SM sphaleron transitions.

\section{Sterile neutrino phenomenology}
The Majorana mass matrix $M$ in Eq.~(\ref{eq:Lagrangian}) sets the value of the new physics energy scale: as long as its dynamical origin is not specified, phenomenological constraints from neutrino physics impose loosely bounds on its value, and RH-neutrinos with masses between the MeV and the GUT scale can equivalently account for the observed neutrino masses and lepton mixing. On the other hand, the testability of the hypothesis in Eq.~(\ref{eq:Lagrangian}) strongly depends on the value of $M$: after electroweak symmetry breaking, the mass spectrum of the model is generally characterised by 3 light (mostly active) neutrinos with masses at the scale $m_\nu$ in Eq.~(\ref{eq:light_masses}), and by $n$ heavy (mostly sterile) neutrinos with masses at the scale $M$ (up to $\mathcal{O}(\Theta^2)$ corrections); the coupling of the sterile neutrinos with the SM gauge bosons is suppressed with respect to the active ones by a factor of order $\Theta$.
If a sterile neutrino is lighter than the $W$ boson, its decays are mediated by off-shell gauge bosons, making it relatively long-lived, while its mass scale is accessible for production in current collider experiments. Such a particle can be searched for by looking, for instance, for displaced vertices in LHC events.

Concerning the early Universe evolution, RH-neutrinos in this mass range tend to equilibrate at late times, around the electroweak phase transition temperature, $T_\text{EW} \approx 140$ GeV, due to their feeble Yukawa couplings dictated by the relation~(\ref{eq:light_masses}). While they are generated from the thermal plasma, their CP-violating scatterings, oscillations and decays generate a lepton asymmetry; sphaleron transitions are only effective at temperatures above $T_\text{EW}$, so that the washout processes due to the late equilibration of the RH-neutrinos do not affect the generated BAU below $T_\text{EW}$. This is the so-called freeze-in (or low scale) leptogenesis mechanism~\cite{freeze-in_lepto}.

\section{Low-scale leptogenesis and testability}
The phenomenology of sterile neutrinos at collider experiments primarily depends on two parameters: their masses $M_i$ and their active-sterile mixing $\mathcal{U}_{\alpha i}$ with the active flavour $\alpha$. For masses below the electroweak scale and in the range of mixing values allowed by experimental constraints, these states are relatively long-lived, and can travel for observable macroscopic displacements before decaying~\cite{Drewes:2019fou}.

The minimal realisation of the Lagrangian~(\ref{eq:Lagrangian}) accounting for neutrino oscillation data requires $n=2$ RH-neutrinos~\cite{SM+2RHN}: the resulting model exhibits a constrained flavour pattern, in which the ratios of couplings to different SM flavours, $ \mathcal{U}_{\alpha i}/\mathcal{U}_{\beta i}$, are bounded and no large hierarchies in the flavour structure are allowed~\cite{flavour_2RHN}. As a result, there exist an upper bound on the active-sterile mixing for which leptogenesis is viable~\cite{lepto_2RHN}: the reason is that too large active-sterile couplings enable the early equilibration of RH-neutrinos and thus a too large asymmetry washout. The washout is a flavour-dependent process, thus it is in principle possible to store the generated asymmetry in a feebly coupled flavour where it is protected from washout, while the other mixings are kept large allowing for collider testability. However, the ``democratic'' flavour structure of the $n=2$ model implies that the upper bound in the region viable for leptogenesis is smaller than current experimental upper bounds on the active-sterile neutrino mixings.

The situation is different in the model with $n=3$, where a RH-neutrino is added for each SM LH one: in this scenario the viable region for leptogenesis is enlarged~\cite{Abada:2018oly}. An obvious reason for this is the extended number of free parameters, which for instance allows for very large hierarchies in the flavour couplings. But, beyond that, there exist  dynamical mechanisms which are peculiar to the $n=3$ scenario - and that are not present in the $n=2$ case - that enhance the lepton asymmetry production. For instance, a new asymmetry source term (absent in the case of $n=2$) is present  in the evolution equations; moreover, the interplay of 3 RH-neutrinos can give rise to the resonant production of a lepton asymmetry. Remarkably, the structure of the $M$ and $F$ matrices (see Eq.~(\ref{eq:Lagrangian})) leading to a dynamical resonant enhancement of the lepton asymmetry can result as a direct consequence of an underlying $B-\bar{L}$ approximate symmetry (difference between baryon $B$ and a generalised lepton number~\cite{Abada:2018oly} $\bar{L}$), which in turn allows for solutions featuring sizeable active-sterile mixing values and no fine-tuning, thus increasing the testability perspectives~\footnote{Further aspects of the connection between an approximate $B-\bar{L}$ symmetry and freeze-in leptogenesis have been also previously addressed~\cite{L_number-lepto}.}.

\begin{figure}
\begin{minipage}{0.5\linewidth}
\centerline{\includegraphics[width=1\linewidth,
]{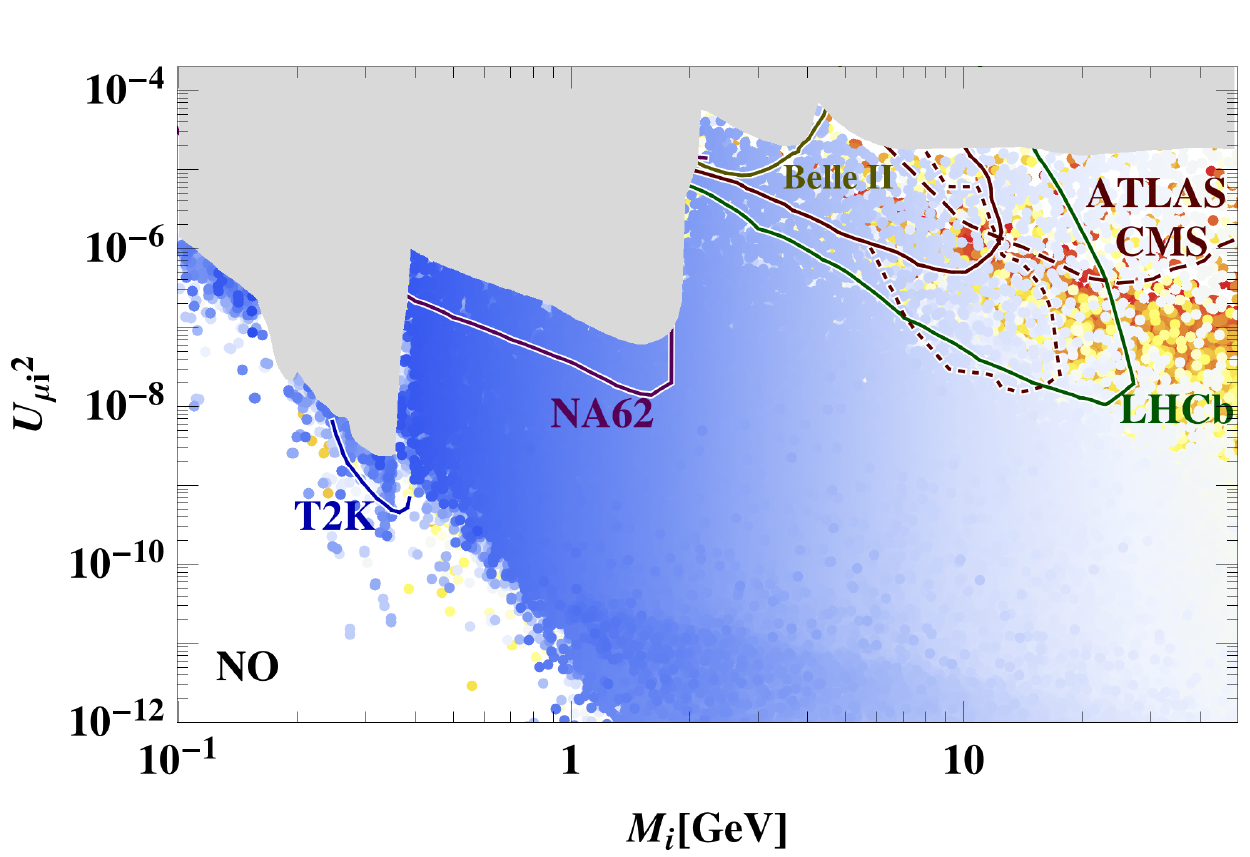}}
\end{minipage}
\hfill
\begin{minipage}{0.5\linewidth}
\centerline{\includegraphics[width=1\linewidth,
]{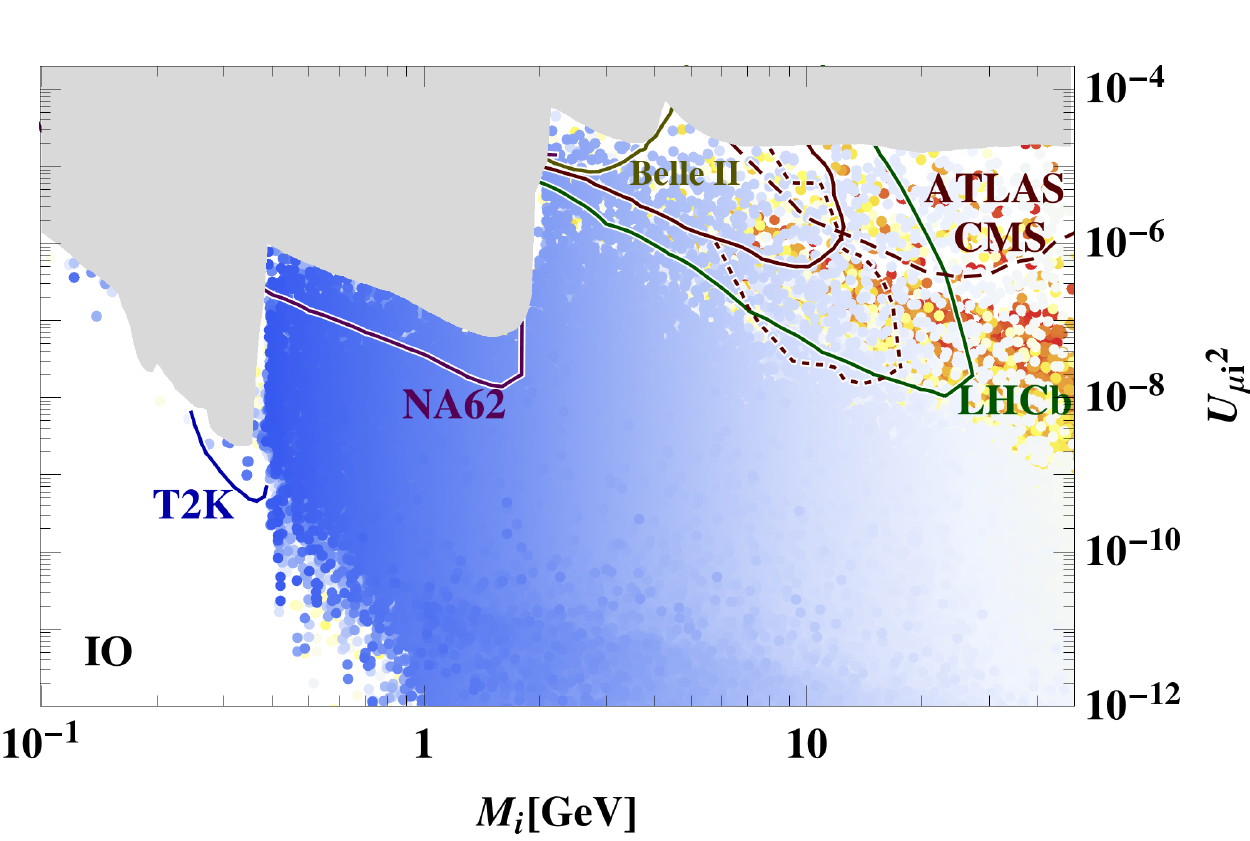}}
\end{minipage}
\caption{Active-sterile mixing in the muon flavour $U_{\mu i}^2$, as a function of the sterile neutrino mass $M_i$, for viable solutions accounting for neutrino masses and mixings as well as BAU, for normal (left) and inverted (right) ordering of the active neutrino mass spectrum. (Not) fine-tuned solutions are represented in (blue) red, see~\protect\cite{Abada:2018oly} for details.  The grey area is excluded by direct experimental searches of sterile neutrinos, while the lines represent the expected sensitivities of ongoing experiments; LHC experiments sensitivities  are based on $pp$ searches.}
\label{fig:3RH-testability}
\end{figure}

As a consequence of the above described phenomenology, the  parameter space for low-scale leptogenesis is extended in the $n=3$ scenario: as reported in Fig.~\ref{fig:3RH-testability}  the upper bound on the viable active-sterile mixing simply results from the current experimental bounds from laboratory searches. This implies that any future improvement on the current experimental constraints in the mass range [0.1, 50] GeV will also translate into an experimental test of the leptogenesis hypothesis in the SM extended with 3 RH-neutrinos.

\section{Testing leptogenesis with heavy ion collisions at the LHC}
The sensitivity curves for LHC experiments reported in Fig.~\ref{fig:3RH-testability} assume standard proton-proton ($pp$) collisions. Sterile neutrinos with masses below the EW scale are relatively light particles compared to LHC energies, but they generally feature feeble couplings. We thus expect the experimental sensitivity to this kind of particle to be more strongly related to the number of accessible events (intensity frontier) than to the available collision energy (energy frontier). In this perspective, an interesting channel to look for sterile neutrinos at the GeV scale is represented by the collision of heavy ions at the LHC. Originally designed to study the dynamics of the quark-gluon plasma (QGP), heavy ion runs can as well test the new physics hypothesis~\cite{Bruce:2018yzs},  profiting from new production mechanisms induced by  strong electromagnetic interactions or by thermal processes in the QGP, or from the enhanced combinatorics in the number of parton level interactions. The latter feature can be advantageous in the search of feebly coupled sterile neutrinos: each nucleus ${}^A_Z N$ indeed contains $A$ nucleons, and in central $NN$ collisions the number of parton level interactions is enhanced by a factor $\approx A^2$, which can amount to four orders of magnitude in the case of the lead isotope ${}^{208}_{\phantom{2}82}$Pb currently used at LHC. Long-lived and feebly interacting sterile neutrinos produced in central collisions easily escape the noisy QGP environment, which only extends to few femtometers. They then decay (semi-)leptonically, and can be searched for by looking for events with displaced vertex signatures.

Heavy ion runs at LHC have both advantages and drawbacks compared to proton runs, and their reach depends on the model under consideration~\cite{ref:HIatLHC}. They feature a smaller collision energy, due to the smaller charge to mass ratio compared to protons, and can deliver only a much lower instantaneous luminosity, due to machine limitations and a faster decay of the beam intensity. On the other hand the parton-level cross section is  enhanced by a factor $\approx A^2$ in central collisions, and the primary vertex mis-identification is negligible in heavy ion runs, due to the absence of pile-up. Finally, although heavy ion collisions are in general characterised by a larger multiplicity of charged tracks per bunch crossing compared to protons, the difference is expected to amount at most to a factor of 2 in the High-Luminosity LHC era, due to the large pile-up the experiments will have to face in this LHC configuration for $pp$ collisions.

\begin{figure}
\begin{minipage}{0.5\linewidth}
\centerline{\includegraphics[width=1\linewidth,
]{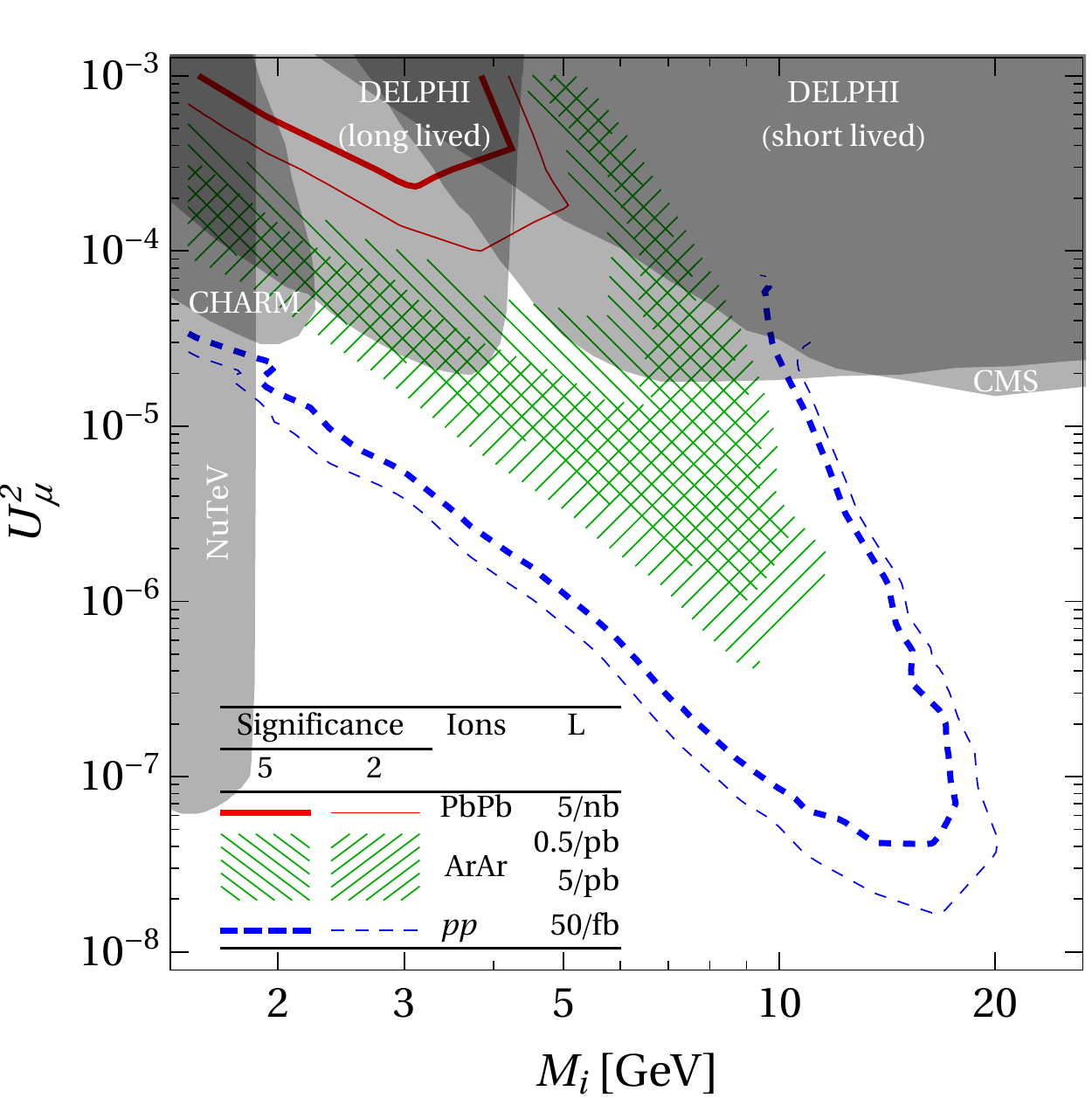}}
\end{minipage}
\hfill
\begin{minipage}{0.5\linewidth}
\centerline{\includegraphics[width=1\linewidth,
]{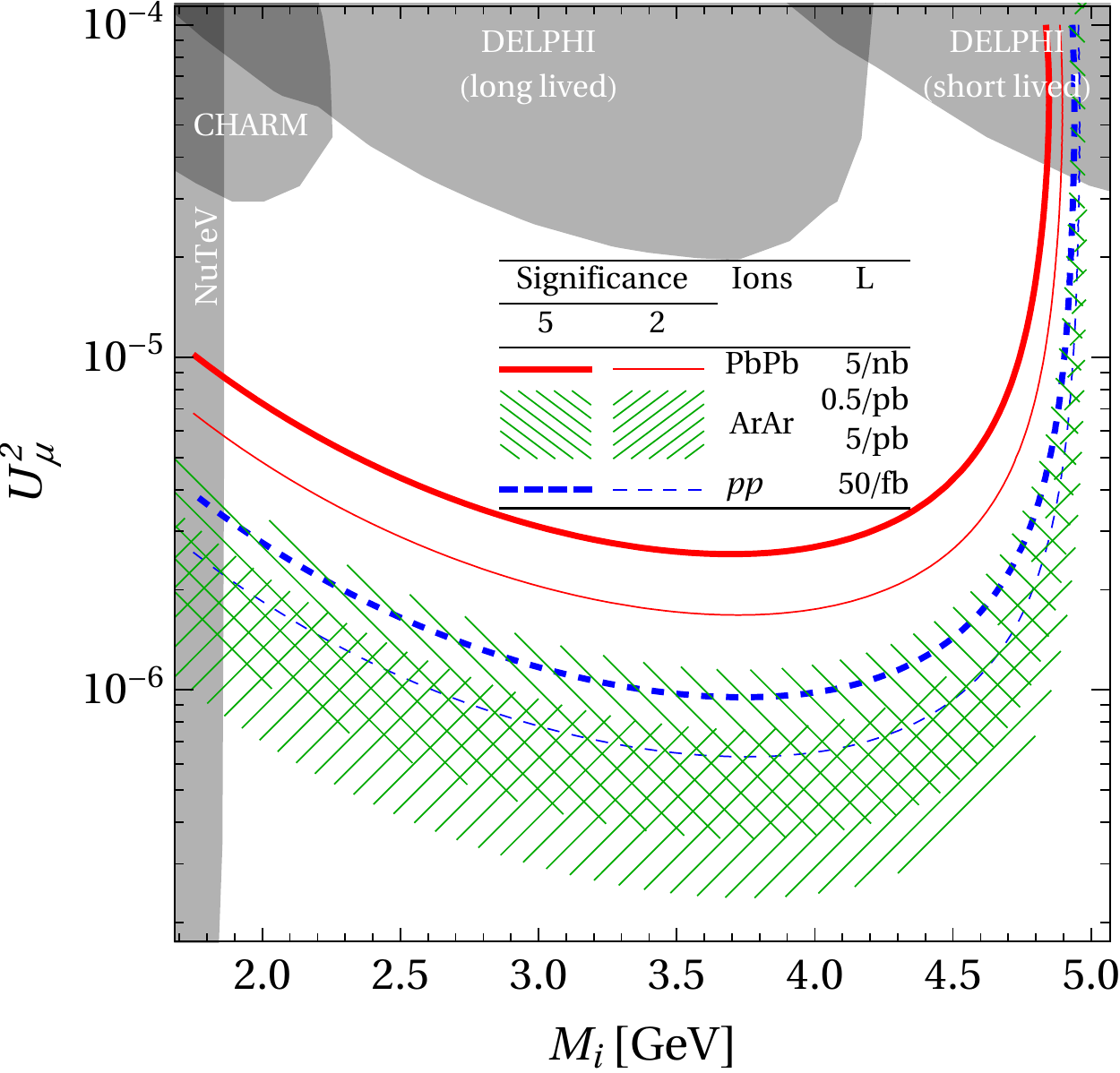}}
\end{minipage}
\caption{Sensitivity of the CMS detector to heavy neutrinos produced in the decay of $W$ bosons (left) and $B$ mesons (right), from the collision of protons (dashed blue lines), Ar (hashed green bands) and Pb (continuous red lines) with integrated luminosities corresponding roughly to one month of running. The green band takes into account the current uncertainties on the performance of LHC with Ar isotopes, which affect the achievable beam intensity.}
\label{fig:HI-sensitivity}
\end{figure}

The lower instantaneous luminosity that characterises heavy ion runs can actually be advantageous for certain searches, since the smaller event rate enables the ATLAS and CMS experiments to significantly lower their trigger thresholds~\cite{ref:HIatLHC}; this allows e.g. to search for signatures with low transverse momentum, that characterise scenarios involving light mediators.
We quantify the improvement in sensitivity that can be achieved with heavy ion searches by using the SM extended with RH-neutrinos as a benchmark model, and considering two different production mechanisms, namely sterile neutrinos produced in the decay of a $W$ boson and of a $B$ meson. In our simulation, we look for signatures triggered by a primary muon produced in the $W$ or $B$ decay (where the sterile neutrino is concurrently produced), and search for displaced muons from secondary vertices (where the sterile neutrino decays) with a minimum displacement of 5 mm. For the $pp$ analysis we adopt a realistic CMS online trigger value on the transverse momentum of the first muon, asking for $p_T > 25$ GeV, while for the heavy ion analysis we adopt a lower trigger threshold, $p_T > 3$ GeV, roughly corresponding to the minimal transverse momentum allowing a charged particle to cross the CMS muon chambers. The results of the analysis are reported in Fig.~\ref{fig:HI-sensitivity}, for equal running time of protons, Pb and Ar isotopes: for the $W$-mediated production, a large fraction of events  has a transverse momentum $p_T$ larger than 25 GeV, due to the large mass of the $W$ boson. Thus, there is not a significant improvement in lowering the trigger threshold, and proton collisions result to be more competitive. Nevertheless, relatively light ions such as Ar can allow to test the production and dynamics of sterile neutrinos in a complementary environment as it is the QGP. The situation is very different for $B$ meson-mediated sterile neutrinos: in such a case, due to the relatively small mass of the $B$ meson, the majority of muons are produced with a transverse momentum smaller than 25 GeV. Lowering the trigger threshold thus allows to access an entirely new kinematical region, and Ar runs are in general more competitive than proton ones for this kind of searches~\cite{ref:HIatLHC}.

\section{Conclusion}
The extension of the SM field content by the addition of $n$ RH-neutrinos can simultaneously account for the massive nature of active neutrinos and for the observed value of the BAU. If the RH-neutrino mass scale lies below the electroweak one, these particles can produce the BAU via freeze-in leptogenesis, and can be searched for in collider experiments by looking for displaced vertex signatures. In the scenario with $n=2$ there exists an upper bound on the value of the active-sterile mixing for which leptogenesis is viable, while for $n=3$ the upper bound is determined by current experimental constraints.

Sterile neutrinos with masses at the GeV scale can be produced in $B$ meson decays: in this case their signatures at collider are characterised by a small transverse momentum, generally smaller than the standard online trigger thresholds on $p_T$. By using heavy ions at the LHC it is possible to significantly lower the trigger threshold, thanks to the lower instantaneous luminosity that characterises heavy ion runs, while at the same time profiting from the parton-level cross section enhancement $\approx A^2$ due to the $A$ nucleons present in each nucleus. Lowering the trigger threshold allows to access an entirely new kinematical region: as a result, intermediate mass ions such as Ar provide, in the search for sterile neutrinos produced in $B$ meson decays, a better sensitivity per unit of running time compared to protons.

\section*{Acknowledgments}
The results presented in this document have been derived in collaboration with A. Abada, G. Arcadi, V. Domcke, M. Drewes, A. Giammanco, J. Hajer, J. Klaric and O. Mattelaer.
This project has received funding from the European Union’s Horizon 2020 research and innovation programme under the Marie Sklodowska-Curie grant agreement No 750627.

\end{document}